\documentclass[12pt,preprint]{aastex}

\usepackage{graphicx}

\shorttitle{Decline of V603 Aql}
\shortauthors{Johnson et al.}

\begin{document}

\title{Nova Aquilae 1918 (V603 Aql) Faded by 0.44 mag/century from 1938-2013}

\author{Christopher B. Johnson and Bradley E. Schaefer}
\affil{Louisiana State University, Dept. of Physics and Astronomy, Baton Rouge, LA, 70802}

\author{Peter Kroll}
\affil{Sonneberg Observatory, D 96515 Sonneberg, Germany}

\author{Arne A. Henden}
\affil{American Association of Variable Star Observers, 49 Bay State Road, Cambridge MA 02138}


\begin{abstract}
We present the light curve of the old nova V603 Aql (Nova Aql 1918) from 1898-1918 and 1934-2013 using 22,721 archival magnitudes. All of our magnitudes are either in, or accurately transformed into, the Johnson $B$ and $V$ magnitude systems. This is vital because offsets in old sequences and the visual-to-$V$ transformation make for errors from 0.1-1.0 magnitude if not corrected.  Our V603 Aql light curve is the first time that this has been done for any nova.  Our goal was to see the evolution of the mass accretion rate on the century time scale, and to test the long-standing prediction of the Hibernation model that old novae should be fading significantly in the century after their eruption is long over.  The 1918 nova eruption was completely finished by 1938 when the nova decline stopped, and when the star had faded to fainter than its pre-nova brightness of $B=11.43 \pm 0.03$ mag.  We find that the nova light from 1938-2013 was significantly fading, with this being seen consistently in three independent data sets (the Sonneberg plates in $B$, the AAVSO $V$ light curve, and the non-AAVSO $V$ light curve).  We find that V603 Aql is declining in brightness at an average rate of $0.44 \pm 0.04$ mag per century since 1938.  This work provides remarkable confirmation of an important prediction of the Hibernation model.
\end{abstract}

\section{Introduction}

Cataclysmic variables (CVs) are short-period, semi-detached binary systems which consist of a white dwarf (WD) accreting matter from a low mass companion \citep{warn95}.  CVs come in many classes, with the two major classes being nova (thermonuclear runaway explosions on the surface of the white dwarf, see Bode \& Evans 2008) and dwarf novae (with instabilities in the accretion disk, see Bath 1975).  Our community has a weak but widespread conclusion that any single CV system must somehow transform itself between classes in some cyclic manner.  Here are three points of evidence: (1) Dwarf novae are slowly accumulating material onto the white dwarf and must eventually become a nova.  Dwarf novae have a systematically lower accretion rate than novae, and so dwarf novae must have some means to greatly increase their accretion rate so as to turn into systems that we see produce novae.  (2) There are greatly too few old novae in the sky to account for the observed nova rate, so old novae must somehow be able to change their appearance, for example by secular decreases in their accretion rate and turning into dwarf novae or perhaps even becoming detached \citep{pat84}.  (3) Shara et al. (2007; 2012) have found two dwarf novae that have old nova shells centered on them, proving that at least these two low-accretion-rate systems were both high-accretion-rate novae in the past.

General evolution of CVs has become one of the major problems in the field, and a variety of models have been proposed.  The first was the Hibernation model proposed by \citet{shar86}, where nova evolution is driven by the inevitable slight disconnect forced on the binary when the system loses mass during a nova event.  The Hibernation hypothesis has long been controversial, with few in the community advocating the full model.  The basic mechanism works, so the question is how deep the hibernation goes. Alternative models (e.g., Vogt 1989; Patterson et al. 2013) also predict the steady fading of old novae by the same mechanism.

The long time scale evolution of novae has never been really tested.  The reason is that the evolution happens on time scales of centuries to millennia, and such data is impossible to get.  If we look on the time scale of nearly a century, the expected effects are not large.  For example, Hibernation predicts a decline rate that should average to perhaps 1 magnitude per century, although the uncertainties on the rate of fading are substantial (Shara et al. 1986; Kovetz et al. 1988).  All pre-1970's magnitudes have systematic errors of up to one magnitude and more (Figures 3 and 5 of Sandage 2001), so it becomes impossible to test old nova evolution unless these effects are corrected.  Another technical problem is that many of the old magnitudes are visual and there are systematic color terms that are required to transform to the modern $V$ magnitude system.  To be complete, we should point out that two prior attempts have been made to measure the fading of old novae. The attempt by \citet{vog90} failed due to a poor method based on the nova amplitude, and the attempt of Duerbeck (1992) failed due to the lack of correction for visual magnitudes or correction for erroneous old sequences.  The requirement to get nearly-century old data also has problems because the number of published magnitudes is fairly small. With horrible time coverage, the trends are hard to see above the usual flickering inherent in the light curve of all CVs.  So the reason for the lack of any viable prior tests is due to the two technical problems (bad old sequences and mixing visual and $V$ magnitudes), while published magnitudes are always inadequate.

In this paper, we solve the dual problem of not having much old data and not having these magnitudes on the modern magnitude system.  First, we get large amounts of unpublished data from archival plates reaching back to the 1800's, and from the large data base of magnitudes going back to 1918 in the archives of the {\it American Association of Variable Star Observers} (AAVSO).  Second, by getting the original sequences of comparison stars as used by the original observers, we can accurately correct the old reported magnitudes into the modern Johnson $B$ and $V$ magnitude systems.  In this paper, we apply these two solutions to the brightest of all  known novae, Nova Aquilae 1918 (V603 Aql).  Our goal is to use this as a test case to answer the question of whether old novae are fading as predicted.

\section{V603 Aql Light Curve} 

The 1918 nova eruption produced the all-time brightest known nova event, peaking at -0.5 mag. The eruption light curve has been previously published in a variety of places (e.g., Campbell 1919; Payne-Gaposchkin 1964), but none of these follow the nova until it is back to quiescence.  Strope et al. (2010) have produced a comprehensive eruption light curve. They classify the light curve as O(12), indicating that it had oscillations superposed on the usual smooth decline and that the time it took to decline by three magnitudes from the peak (i.e., $t_3$) was 12 days. \citet{str10} measured that the eruption light curve took 6800 days (i.e., 18.6 years, until 1937) to return to quiescence by means of linear extrapolation of the fitted light curve (with logarithmic time) to the steady post-eruption level.

\subsection{Pre-eruption Light Curve}

We have measured the pre-eruption $B$ magnitude for V603 Aql with the archival plates at the Harvard College Observatory.  These plates are all in the $B$ magnitude system (they were the original source for the definition of the system) with an essentially zero color term.  The comparison stars that we used were all with modern $B$-band magnitudes, so the derived magnitudes for V603 Aql are all exactly in the Johnson $B$ magnitude system.  With much experience of experimental measures, the one-sigma measurement uncertainty is 0.15 mag on average for these plates.  Our 50 pre-eruption magnitudes going from 1898 to 20 days before eruption are presented in Table 1. The pre-eruption magnitudes have an RMS scatter of 0.23 mag, which is likely larger than our typical measurement uncertainty due to usual flickering. The average pre-eruption magnitude is 11.43 $\pm$ 0.03 mag.

\subsection{Post-Eruption Light Curve}

We have constructed the post-eruption light curve primarily from two sources: the Sonneberg archival plates from 1934 to 2004 for the $B$-band and the AAVSO light curve from 1934 to 2013 for the $V$-band.  Full details of our extraction of magnitudes from the Sonneberg plates has been covered in Collazzi et al. (2009).  Just as with the Harvard plates, the Sonneberg plates all have near-zero color terms. Modern Johnson $B$ magnitudes were used for the comparison stars, so the resultant magnitudes for V603 Aql are all correctly calibrated to the Johnson $B$ magnitude system. We used visual magnitudes made by Steavenson between 1928-1953 (Steavenson 1928, 1934, 1935, 1936, 1938, 1939, 1947, 1948, 1950, 1953).  Also, we have $V$-band observations from \citet{lan68}, \citet{lan73}, \citet{bru80}, \citet{bru91b}, \citet{bru91c}, \citet{pat97}, \citet{sher83}, \citet{wal57}, and \citet{sul04}, with all of these being on the Johnson $V$ magnitude system.  Additional B-band magnitudes taken with CCD cameras appear in the AAVSO database, all of which are already on the Johnson $B$ magnitude system.

\subsection{Transformation from Visual to $V$ magnitudes} 

All of our magnitudes from plates and CCDs are already in the Johnson $B$ and $V$ magnitude systems.  But all of our visual data is only {\it close} to the Johnson $V$ magnitudes.  Typical differences between $V$ and visual magnitudes amount to a tenth of a magnitude, mainly because V603 Aql is very blue compared to most of its comparison stars.  A further problem is that the old-time comparison star magnitudes are imperfect, with typical errors of a third of a magnitude in this case.  This translates into systematic errors for all early observations made with the charts, while the chart errors vary with time.  Therefore, all the visual measures must be transformed to the Johnson $V$ magnitude system.

The human spectral sensitivity differs from that of the Johnson $V$ system, with the human eyes having more blue sensitivity.  The transformation from visual magnitudes ($m$) to $V$-band magnitudes ($V$) has been exhaustively measured for many observers as reported in \citet{stan99}.  They find 
\begin{equation}
m = V + 0.21\times(B-V)
\end{equation} 
over a very wide range of $B-V$ colors.  The $B$ and $V$ magnitudes for the comparison stars are now well measured, with compilations in the AAVSO Photometric All Sky Survey (APASS).  With this, we can calculate the visual magnitude of each comparison star.  V603 Aql has an average color of $B-V$= -0.04 \citep{bru94}, which allows us to transform visual-to-$V$ for the old nova.

The correction for the old charts is straightforward.  Each star in the comparison sequence has a modern $B$ and $V$ magnitude, so we can calculate each visual magnitude, $m$.  Each star also has a given magnitude recorded on the chart, which we label as $\mu$.  Essentially all visual observations are made with the observer judging the apparent visual magnitude of the target against two nearby comparison stars that bracket the target in brightness and are the closest in brightness to the target.  We will describe the magnitudes of the just-brighter comparison star with the subscript `$b$' and the just-fainter comparison star with the subscript `$f$'.  The observer makes a judgement as to the fraction, $F$, of the brightness difference that the target star is from the brighter star.  Thus, $F=(m-m_b)/(m_f-m_b)$.  The magnitude {\it reported} by the observer, $\mu$, will be the observed fraction of the difference between the two comparison stars as calculated from their charted magnitudes, so we also have $F=(\mu - \mu_b)/(\mu_f - \mu_b)$.  Equating these two expressions for $F$, and we are left with one equation involving the quantity that we want ($m$, soon to be converted to $V$) and known quantities.  So, 
\begin{equation}
V=m_b + [(m_f-m_b)(\mu - \mu_b)/(\mu_f - \mu_b)] -0.21\times(B-V).
\end{equation} 
The uncertainty in this transformation is small (Stanton 1999), so the dominant uncertainty in $V$ is simply the usual measurement error which is  $\pm$ 0.20 mag.  So now we have a full and accurate prescription for transforming from visual to $V$.

Old magnitudes from the literature almost always report their comparison stars and their adopted magnitudes.  So, for example, Steavenson's magnitudes were made with a sequence reported in \citet{ste38}.  For the AAVSO observations, we have gone through the archives at the AAVSO Headquarters to pull out the full set of charts as used by actual observers.  The chart used from 1918 until 1986 is that of 184300(d) as based on Harvard photometry.  The 1986 chart was used until late 2008, when a new chart (the current chart) was introduced. The modern Johnson $B$ and $V$ magnitudes for all sequence stars come from the APASS photometry \citep{hen12}. 

\subsection{Final Light Curve}

We have collected 22,721 Johnson $B$, $V$, and visual magnitudes into Table 1 (full table available upon request).  We have 50 pre-eruption $B$ magnitudes from Harvard (1898-1918), 538 post-eruption $B$ magnitudes from Sonneberg (1928-2004), 15 $B$ magnitudes from the AAVSO (2010-2012), 2 $B$ magnitudes from AFOEV (2012), 18,980 visual magnitudes from the AAVSO (1934-2013), 61 visual magnitudes from Steavenson (1928-1953), and 3,075 $V$ magnitudes from many literature sources (1957-2013).  For most magnitudes, the individual error bars were not measured, so we have used typical values of $\pm$ 0.15 for the Harvard and Sonneberg plates, $\pm$ 0.20 for visual observers, and $\pm$ 0.02 for CCD measures.  

The `large' error bars for non-CCD magnitudes is irrelevant because the ordinary flickering on all time scales of V603 Aql is always larger.  That is, the $\pm$ 0.20 mag uncertainty from visual observers contributes only a fraction of extra variance to our light curve because a much larger variance arises from flickering in the star.  The 0.15-0.20 mag measurement errors for single observations is also irrelevant because we always have a large number of observations ($N$), so the measurement error is $0.20/ \sqrt{N}$ and is always small.

For long, time series observations within one night, we made nightly averages and use this average to represent each nightly brightness.  

With the ubiquitous flickering of V603 Aql, our light curve plot shows the usual scatter of points, with the true behavior partly hidden by all the overlapping ink for all the points.  To better show the long-term behavior, we have binned our light curve into yearly averages for each calendar year from 1934.  With the large number of points per year, we can directly determine the measurement error bar of the average as the larger of RMS/$\sqrt{N}$ or $0.15/ \sqrt{N}$, where $N$ is the number of included observations and RMS is their scatter. For bins with hundreds of magnitudes, this leads to unrealistically small error bars because the measurement errors will be dominated by systematic errors.  We have put a lower limit on the error bars of 0.02 mag.  The light curve is displayed in Figure 1.

\section{V603 Aql Is Fading} 

The Hibernation model predicts that old novae will be fading at roughly the rate of one magnitude per century.  Other models (e.g., Patterson 2012) also predict fading long after the eruption is completely over and all the eruption-related transients have completely died down.  The whole purpose of our study of V603 Aql is to test these models of nova evolution.  A glance at Figure 1 shows that V603 Aql is fading, and the rate is roughly half a magnitude per century.  

The steeper decline in $V$ before 1938 is likely just the tail end of the nova eruption.  To avoid confusing the end of the eruption with any fading due to evolution, we must only look for a trend after the eruption is over.  Strope et al. (2010) fitted the fading tail and found that it reached quiescence in early 1937 (i.e., 6800 days after peak).  A second convincing means to determine the end of the eruption is to see when the nova has faded to the pre-eruption level ($B=11.43\pm0.03$), which is certainly true by 1938. Thus, we confidently take 1938 to be the year by which the eruption is completely over. 

After 1938, there are some fluctuations in this otherwise steady decline.  The correlation between the variations in the $B$ and $V$ light curves is vague and unconvincing.  From 1975 to 1981, both colors show a decline compared to the best fit linear trend, but the $B$-band light curve immediately brightens from 1981 to 1990 while the $V$-band light curve holds constant over the same interval. Neither the structure in the $V$ light curve from the 1960's, nor the outlier 1964 $B$ magnitude are shown in the other color.  V603 Aql has never been seen to change its color significantly (Bruch \& Engel 1994) or systematically in any substantial manner, so the deviations between the $B$ and $V$ light curves are likely caused by some sort of unrecognized systematic error.  This alerts us that our light curve still has systematic variations at the 0.1 mag level.  Fortunately, such problems are small compared to the observed fitted decline of $\sim$0.4 mag over 76 years.  

Another way to get an idea as to systematic errors is to compare the AAVSO $V$ light curve with the non-AAVSO $V$ light curve.  These are two completely independent measures of the same function, so significant variations can only be ascribed to unknown systematic problems.  A direct comparison is problematic because the non-AAVSO light curve has many gaps.  Nevertheless, a simple fitting of the best line returns identical intercepts and slopes of $0.46 \pm 0.08$ and $0.58 \pm 0.18$ mag per century for the AAVSO and non-AAVSO binned light curves respectively from 1938-2002. This comparison suggests that systematic problems in the derived slope of the light curve are negligibly small.

Whether through intrinsic variations in V603 Aql or due to small residual systematic errors in measuring the light curve, we realize that the 1938-2013 light curve will show deviations from a straight line.  Such deviations will result in a chi-square fit of the light curve to a simple line that has a large reduced chi-square.  A more realistic analysis will allow for some (presumed constant) variations (with RMS scatter of $\sigma_{sys}$) superposed on a steady linear decline.  In essence, we have the total one-sigma variations in the yearly-binned magnitudes equal to $\sqrt{\sigma_{meas}^2+\sigma_{sys}^2}$.  If we set $\sigma_{sys}$ such that the reduced chi-square of the fit is near unity, then we get realistic error bars in the slope.  With this, we set $\sigma_{sys}=0.10$.  


The 1938-2013 $B$ and $V$ yearly-binned light curves for V603 Aql in Figure 1 have fitted slopes of $0.30\pm0.08$ and $0.50\pm0.05$ mag per century respectively. These two slopes are different at the two-sigma level, and we judge them to be the same to within the error bars.  (Such differences are easily caused by, for example, the lack of $B$-band magnitudes from 2005-2009, such that their inclusion would apparently bring the slopes into much closer agreement.)  The existence of similar slopes in completely independent light curves gives us good confidence that V603 Aql is fading.  With V603 Aql observed to be essentially constant in color (Bruch \& Engel 1994), the best measure of the slope is from some combination of the measures in the two colors.  The weighted average of the two slopes is $0.44 \pm 0.04$ mag/century.  

\section{Discussion} 

We began by proposing to test the prediction of nova evolution models that old novae will steadily fade, perhaps at a rate of a magnitude per century, when they are in quiescence, long after the eruption is over.  Any such test requires that we get magnitudes from many decades ago. We realized that technical problems in light curve construction (in particular, the mixing of visual and $V$ magnitudes as well as the ubiquitous and large errors in all early comparison sequences) present insidious systematic errors larger than the effect being sought.  Because of this, there has been no effective prior test of the claim that old novae should be fading as part of CV evolution.  Fortunately, we have also realized that these technical problems can be easily solved (see equations 1 and 2), and so all the old visual magnitudes can be transformed confidently into the modern magnitude scale.  Finally, we have realized that the number of unpublished magnitudes (from the AAVSO and from the Harvard and Sonneberg plates) is an order of magnitude larger than all the published data combined.  With this, we have a simple plan to construct well-sampled $B$ and $V$ light curves over many decades, which can be used to seek the evolution of old novae.

V603 Aql certainly had its eruption completely finished by 1938, and the light curves from 1938-2013 (Figure 1) show a distinct and fairly steady decline.  This decline is seen significantly in three completely independent data sets, so we are confident in its reality.  Our overall average decline rate is $0.44 \pm 0.04$ mag per century from 1938 to 2013.  V603 Aql had already faded to {\it below} its pre-eruption level ($B=11.43 \pm 0.03$) by 1938, so this fading cannot be any sort of a tail to the eruption itself.  Rather the observed fading must be caused by some evolutionary effect, for example, the slight disconnection associated with the mass loss during the nova event as emphasized by the Hibernation model.  Indeed, the first measured decline rate in an old nova reasonably matches the {\it prediction} by the Hibernation hypothesis, and this can be viewed as a major success for the model. Other evolution models predict fading in old novae by the same mechanism as Hibernation, so our evidence does not uniquely point to the Hibernation model.



\acknowledgments{We thank the National Science Foundation for support of this work.}

\clearpage

\begin{deluxetable}{lcccccccccr}
\tablewidth{0pt}
\tablecaption{Individual Magnitudes}
\tablehead{
\colhead{JD}                  && \colhead{Band}     &&
\colhead{m} && \colhead{Source}}
\startdata

2414515.621&&B&&11.2&&Harvard AC29\\
2415546.786&&B&&11.2&&Harvard AC1522\\
2415552.749&&B&&11.4&&Harvard AC1546\\
2415562.744&&B&&11.0&&Harvard AC1571\\
2415966.783&&B&&11.3&&Harvard AM1489\\
2452900.593&&Vis&&11.6&&TDB, AAVSO\\
2452901.325&&Vis&&11.6&&MUY, AAVSO\\
2452901.684&&Vis&&11.8&&HK, AAVSO\\
2452902.569&&Vis&&11.7&&BEB, AAVSO\\
2452907.544&&Vis&&11.4&&BRJ, AAVSO\\
 \enddata
 \tablecomments{The error estimates are described in Section 2.4. The source column contains either the Plate ID or the observer's initials followed by affiliation or a reference. Table 1 is published in its entirety in the electronic edition of the Astrophysical Journal Letters. A portion is shown here for guidance regarding its form and content.}
\end{deluxetable}




\begin{figure}[Ht!] 
\plotone{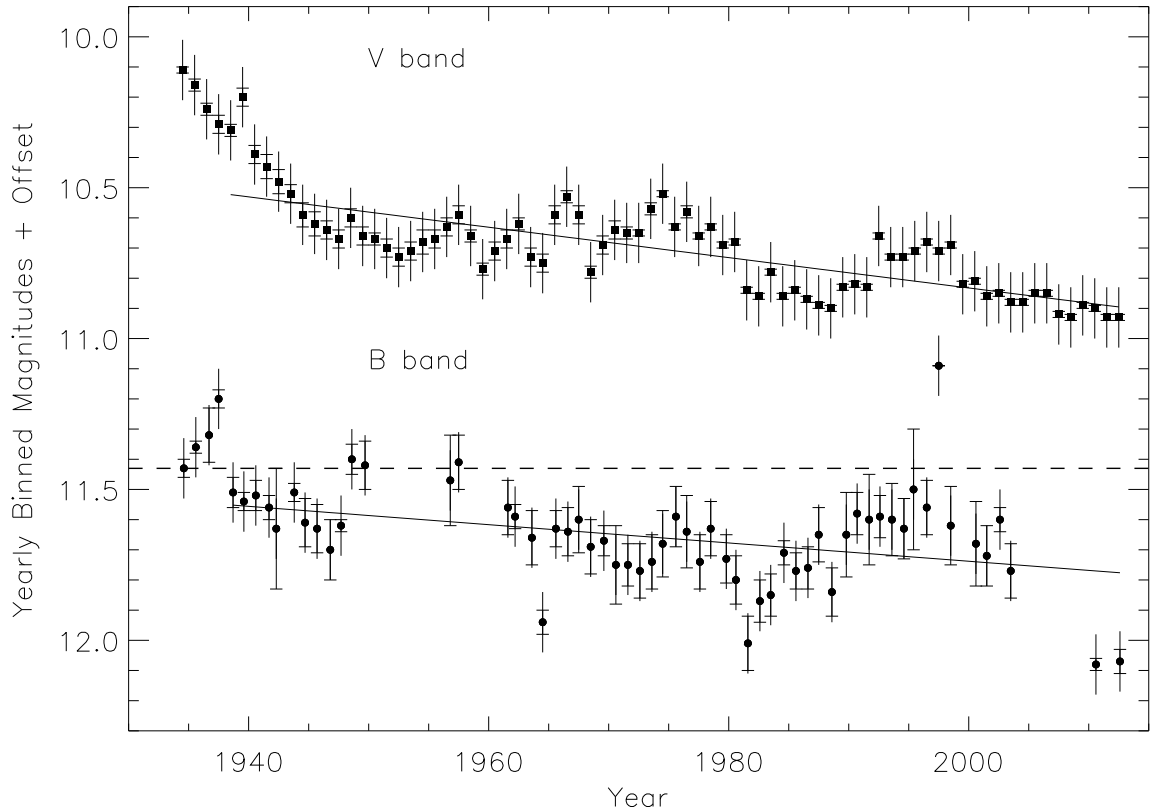}
 \caption{Yearly-binned light curve for V603 Aql.  The bottom curve is the $B$-band yearly-averaged magnitudes from 1934-2013, and is taken primarily from Sonneberg archival plates.  The top curve is for the $V$-band, with an offset upwards of 1.0 mag, taken primarily from the visual magnitudes reported to the AAVSO.  These visual magnitudes have been correctly transformed into Johnson $V$-band measures.  
 Each yearly average has its measurement uncertainty represented by the inner error bars with serifs, while the total uncertainty, including the estimated systematic uncertainty, is represented by the outer error bars.  The {\it pre}-eruption level is at $B=11.43 \pm 0.03$ mag (and so $V=11.47 \pm 0.03$), as represented by the horizontal dashed line, so the eruption is certainly over by 1938.  The point of this figure is that both $B$ and $V$ light curves show V603 Aql to be significantly declining from 1938-2013. A machine-readable table containing the data used to create this figure is available in the online journal.  
 }
 \label{fig:example}
\end{figure}

\end{document}